\documentclass{article}
\usepackage{spconf,amsmath,graphicx}
\usepackage{booktabs}
\usepackage{multirow}
\usepackage{amsmath}
\usepackage{amssymb}
\usepackage{caption}
\usepackage{subcaption}
\usepackage{cite}
\usepackage{comment}
\usepackage{hyperref}

\title{Automated Coronary Arteries Labeling Via \\  Geometric Deep Learning}
\name{Yadan Li$^{1}$, Mohammad Ali Armin$^{3}$, Simon Denman$^{2}$, 
David Ahmedt-Aristizabal$^{2,3}$ }
\address{
$^{1}$ Australian National University, Canberra, Australia. \\
$^{2}$ SAIVT, Queensland University of Technology, Brisbane, Australia. \\
$^{3}$ Imaging and Computer Vision Group, CSIRO Data61, Canberra, Australia. \\
\tt\normalsize david.ahmedtaristizabal@data61.csiro.au
}

\begin{document}
\ninept
\maketitle

\begin{abstract}
Automatic labelling of anatomical structures, such as coronary arteries, is critical for diagnosis, yet existing (non-deep learning) methods are limited by a reliance on prior topological knowledge of the expected tree-like structures. As the structure such vascular systems is often difficult to conceptualize, graph-based representations have become popular due to their ability to capture the geometric and topological properties of the morphology in an orientation-independent and abstract manner.
%
However, graph-based learning for automated labeling of tree-like anatomical structures has received limited attention in the literature. The majority of prior studies have limitations in the entity graph construction, are dependent on topological structures, and have limited accuracy due to the anatomical variability between subjects.
%
In this paper, we propose an intuitive graph representation method, well suited to use with 3D coordinate data obtained from angiography scans. We subsequently seek to analyze subject-specific graphs using geometric deep learning. The proposed models leverage expert annotated labels from 141 patients to learn representations of each coronary segment, while capturing the effects of anatomical variability within the training data.
%
We investigate different variants of so-called message passing neural networks. Through extensive evaluations, our pipeline achieves a promising weighted F1-score of 0.805 for labeling coronary artery (13 classes) for a five-fold cross-validation.
%
Considering the ability of graph models in dealing with irregular data, and their scalability for data segmentation, this work highlights the potential of such methods to provide quantitative evidence to support the decisions of medical experts. 
\end{abstract}

\begin{figure}[!t]
\centering
\includegraphics[width=0.99\linewidth]{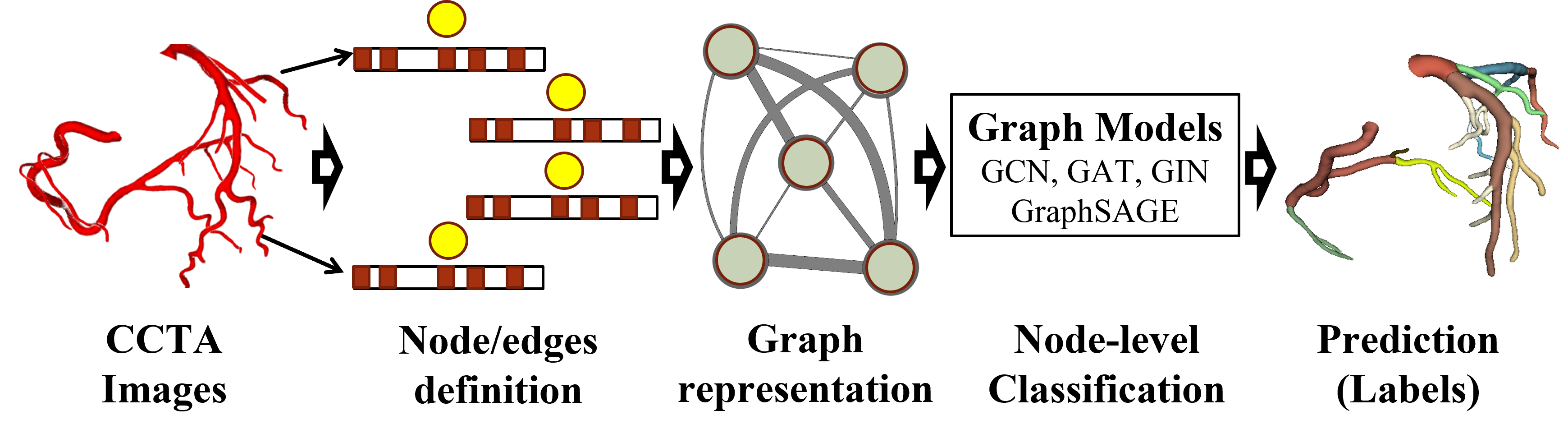}
\vspace{-8pt}
\caption{
Overview of the automated coronary artery labelling framework.
i) Coronary computed tomography angiography (CCTA) images from 141 subjects are annotated into 13 segments (classes) following SCCT guidelines. 
ii) For each CCTA image, a set of coronary artery centerlines for each segment are obtained to define the nodes, edges and node embeddings. Node features comprise spatial features to characterize the entities.
iii) The coronary artery segments and branches for each subject are transformed into a graph.
iv) The graph representation is processed using geometric deep learning models for node-level prediction to label the segments.
}
\label{fig:Fig1}
\vspace{-15pt}
\end{figure}

\vspace{-5pt}
\section{Introduction}
\label{sec:intro}
\vspace{-5pt}

Labelling blood vessels is a critical step in the medical image processing pipeline for the diagnosis of cardiovascular diseases~\cite{chen2002quantitative,arbab2018challenge}.
Unlike spheroid-like organs (\textit{i.e.} the liver or kidney), coronary arteries are divergent, thin, and tenuous. Labeling of the tree-like structures formed by the arteries is complex, with branches of varying sizes and different orientations across individuals and viewpoints. 

Learning from graphs, termed geometric deep learning or graph-based deep learning, is an emerging field that deals with irregularly structured data. Graph learning methods such as Graph Convolutional Networks (GCNs) are powerful tools for the study of non-Euclidean domains, and have been used to support medical diagnosis by representing medical data including brain electrical activity, anatomical structures, and digital pathology, as graphs~\cite{ahmedt2021graph,ahmedt2021survey}.

Traditional non-deep learning methods for coronary artery labeling have used 3-dimensional coronary tree models~\cite{cao2017automatic}. In~\cite{wu2019automated} the authors used bidirectional tree-structured LSTMs, proposing TreeLab-Net to aggregate and pass information through the artery tree. Generally, existing studies achieve high accuracy for the main arteries, but struggle identifying side branches due to large vascular variations and a reliance on simple geometric rules. Furthermore, tree-structured models can only pass messages between nearby layers. 
A graph-based method learns and regresses the locations of arteries directly, and allows for learning of local spatial structures. Graph models can also propagate and exchange local information across the whole image to learn semantic relationships between objects~\cite{tan2021sgnet}.

GCNs were investigated by~\cite{wolterink2019graph} for labeling coronary computed tomography angiography (CCTA) images. In~\cite{wolterink2019graph}, a network with five GCN layers is used to learn local and neighborhood features and optimize the location of nodes in a tubular surface mesh graph. 
Later, the same research group proposed a graph attention network (GAT) for coronary artery segment labeling~\cite{hampe2021graph} of a similar coronary artery tree, which showed improved performance on small leaf branches.
In~\cite{zhang2021corlab}, anatomical and morphological features are used with point cloud deep networks for automatic coronary artery labeling.
A hybrid method is proposed in~\cite{zhou2020hybrid}, where a gated GCN is used to label the anatomical artery tree and logic-based rules are used to refine GCN predictions.
In~\cite{yang2020cpr}, a conditional partial-residual GCN and a hybrid model comprising 3D CNNs followed by bidirectional LSTMS are jointly trained. The hybrid model extracts features along each branch to provide conditions for the graph model~\cite{yang2020cpr}. The two parts are trained end-to-end, capturing both local and global spatial image features.

Geometric deep learning is still in its nascent stages compared to existing deep learning methods for automated labeling. There are challenges associated with the entity graph construction and the complexity of graph training.
%
%
Developing a standardized mechanism to leverage centerline information in graph construction is critical and challenging. Labeling coronary arteries with a GCN requires a fair representation of a segment's location to prevent confusion (\textit{e.g.} changes in the starting point of centerlines across subjects can degrade prediction accuracy). Thus, developing a method to define a graph from centerlines, that is robust and consistent across subjects and datasets, is vital. 

Regarding the complexity of the vascular structures, methods that enable the accurate labeling of main arteries and side branches are required, yet many researchers have observed that proposed GNNs methods perform poorly for small branches, and may even completely omit such structures. This is illustrated by previous studies~\cite{yang2020cpr,hampe2021graph,zhou2020hybrid} failing to consider the left posterior lateral branches (L-PLB) and left posterior descending artery (L-PDA). 

In this paper, we propose a geometric deep learning framework (shown in Fig.~\ref{fig:Fig1}) for coronary artery labeling. Using annotated CCTA images from 141 patients, we first present our method for artery tree graph representation which efficiently defines the origin and last point of centerline, and adopt appropriate embeddings incorporating directional vectors and standard geometric features to represent the characteristics of coronary artery branches.
Next, the generated coronary artery tree is transformed into a linegraph and fed to a graph network for anatomical labeling of 13 artery segments (classes), resulting in a more comprehensive set of labels than earlier efforts 
(9 classes-350 patients~\cite{zhou2020hybrid},
10 classes-71 patients~\cite{hampe2021graph},
11 classes-511 patients~\cite{yang2020cpr}).
We obtain promising results without requiring complex graph models.

The contributions of our work are summarized as follows:
\begin{enumerate}
\vspace{-3pt}
\item We define a standard method for graph representation of coronary arteries. The proposed approach introduces a rule set to construct a graph from the location and direction of artery segments using centerlines. This reduces ambiguity across subjects when mapping coronary artery segments to a graph, and is easily extendable to allow for different node embedding representations.   
\vspace{-3pt}
\item We evaluate different graph neural network methods for labeling coronary artery segments on a privately collected dataset, consisting of CCTA scans from 141 patients manually annotated by three experts. We achieve promising results on this dataset, which includes diversity among subjects and additional complexity from the inclusion of complicated segments and two additional classes (L-PLB, L-PDA) not considered in previous works~\cite{yang2020cpr,zhou2020hybrid,hampe2021graph}.
\end{enumerate}

\vspace{-12pt}
\section{Method}
\vspace{-5pt}

In this section, we detail the graph-based labeling system for challenging and variable anatomical structures such as coronary arteries.
A graph maps regions in the CCTA to nodes in the graph. Then, the labeling problem can be transformed to one of assigning the correct label to each node according (node-level prediction).
A standard entity-graph workflow is developed which consists of the following phases: graph construction (identify nodes, extract embeddings and define edges), and graph modelling (applying deep learning graph models to process the graph). 
For fair comparison, we 
(i) adopt a standard GCN as our graph model to compare node embeddings, since it is currently the most popular model for labelling anatomical structures (vasculature systems and organs).
ii) employ the proposed graph representation and embeddings to compare several graph neural network models, which differ in the way they construct messages, update a node's hidden representation, and obtain the graph-level feature vector, while still following a message passing and readout process.
To obtain node-level predictions, the node embedding is input to a Multi-Layer Perceptron (MLP).

\begin{figure}[!t]
\centering
\includegraphics[width=0.4\linewidth]{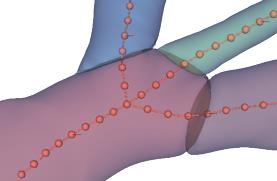}
\vspace{-3pt}
\caption{Illustration of the vessel centerline used to extract features. Each centerline is composed of a list of points obtained via~\cite{antiga2002patient}.  
}
\label{fig:Fig2}
\vspace{-6pt}
\end{figure}

\begin{figure}[!t]
\centering
\includegraphics[width=0.7\linewidth]{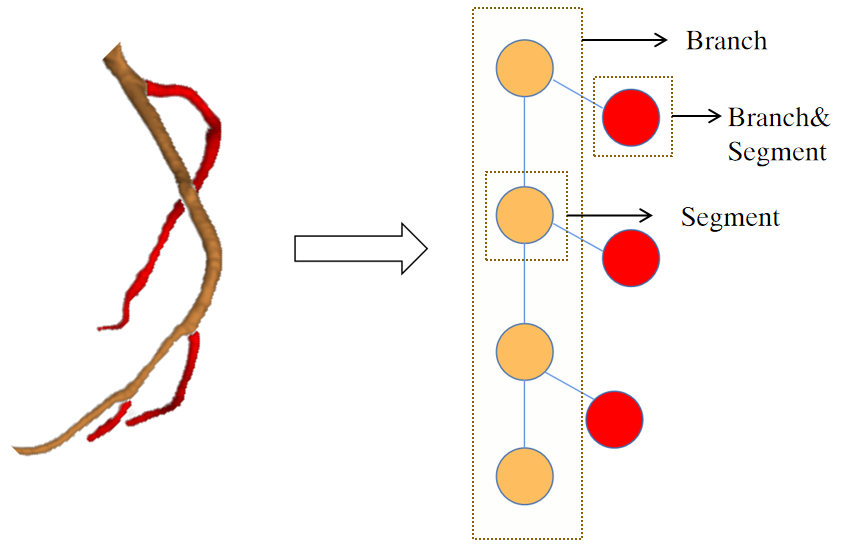}
\vspace{-6pt}
\caption{Graph-based representation of coronary trees segments. 
Segments represent branches after bifurcating. 
Segments are represented by multiple nodes in the graph. 
Visualization of 2 branches of the right coronary arteries: RCA (orange) and AM (red). The same color represents a branch, and nodes correspond to a segment}
\label{fig:Fig3}
\vspace{-10pt}
\end{figure}

\begin{table*}[!t]
\caption{Number of samples of different classes in the dataset.
The reported average, max and min highlight the variability in the dataset.}
\vspace{-5pt}
\centering
\label{table:detail-dataset}
\resizebox{0.98\textwidth}{!}{%
\begin{tabular}{l | c c c c c c c c c| c c c c | c c c c}
\toprule
             & \multicolumn{9}{c|}{\textbf{Left Side}}                         & \multicolumn{4}{c|}{\textbf{Right Side}} & \\
\cline{2-14}  & \textbf{LM} & \textbf{LAD} & \textbf{LCX} & \textbf{R} & \textbf{S} & \textbf{OM} & \textbf{D} & \textbf{L-PLB} & \textbf{L-PDA}       & \textbf{RCA} & \textbf{AM} & \textbf{R-PLB} & \textbf{R-PDA}          & \textbf{Total} & \textbf{Avg} & \textbf{Max} & \textbf{Min} \\ 
\hline
Branches & 136 & 142 & 140 & 64 & 124  & 159 & 182 & 62 & 39 &                   142 &132 & 167 & 102 & 1591 & 11.36 & 24 & 6  \\

Segments & 148 & 502 & 336 & 72 & 194 & 219 & 216 &82  & 41 &                  573 & 321 & 326 & 173 & 3203 & 22.87 & 86 & 15  \\
\bottomrule
\multicolumn{18}{p{550pt}}
{
\textbf{Left side arteries}: left main (LM), left anterior descending (LAD), left circumflex (LCX), Ramus (R), septal artery (S), obtuse margin (OM), diagonal (D), left posterior lateral branches (L-PLB), left posterior descending artery (L-PDA).  
\textbf{Right side arteries}: right coronary artery (RCA), right acute marginal artery (AM), right posterior lateral branches (R-PLB), right posterior descending artery (R-PDA). 
}
\end{tabular}}
\vspace{-10pt}
\end{table*}

\vspace{-8pt}
\subsection{Entity-graph construction}
\label{subsec:egconstruction}
In this section, we present our method for graph construction method, using the initial labeling results. Our coronary artery graph construction approach is composed of three steps: centerline extraction, node and edge definition, and feature extraction.

\textit{Centerline extraction}: 
We use the method of \cite{antiga2002patient} to extract the centerline from the CCTA image. This results in a set of centerlines, each composed of a set of points (see Fig.~\ref{fig:Fig2}). We resample these centerlines to ensure that points are evenly spaced, and adjacent branch starting points are merged to reduce unnecessary segmentation.

\textit{Node and edge definition}:
We adopt similar definitions to~\cite{yang2020cpr}, and whenever branches bifurcate, we treat the resultant segment as a new node. As such, one branch may be represented by multiple nodes in the graph. Here, segments represent branches after bifurcating. This representation is illustrated in Fig.~\ref{fig:Fig3}

To standardize changes in the centerline points in the Cartesian coordinate system, the author in \cite{wu2019automated} proposes a spherical coordinate transform (SCT$S^2$) to convert 3D position coordinates into azimuth and elevation. Due to the phase and other differences in each CCTA image, to obtain the azimuth and elevation, the same $x$,$y$,$z$ axes must be defined for each image. In \cite{wu2019automated}, the center point of the bounding box enclosing all centerline points is taken as the origin, and the x-axis and z-axis of the Cartesian coordinate system are the North Pole and the primary meridian of the spherical coordinate system.

Due to the periodicity of the angle, the range of the angle is limited to $[0,2\pi)$. To overcome instability due to the periodicity, the authors of ~\cite{yang2020cpr} propose using an $S^2$ manifold to represent the angles. This form of spherical coordinate transformation is called SCT$S^2$, and we use a similar transformation in our approach. 

In \cite{yang2020cpr}, the authors select the first control point of each branch as the origin, the $z$ axis is defined as the direction from the first point to the second point, and the vector pointing from first to the last point of centerline lies in the $y-z$ plane. This definition of the origin and last point may be ambiguous when representing the global information of the coronary artery as graph for each subject. As such, to ensure consistency we consider the first point of the first centerline of the left branch to be the origin (generally the centerline of left main), and the definition of the $z$-axis remains unchanged. We define the last point of the last centerline of the right branch (right coronary artery or right posterior descending artery) as the control point which defines the $y-z$ plane.

\textit{Node embeddings}:
In \cite{yang2020cpr} and \cite{wu2019automated}, the authors use similar features based on: i) the direction vector from the first point to the last point of a segment and the tangential vector to the starting point in both 3D and $S^2$; and ii) the first point, center point, and last point of a segment in normalized 3D space and in spherical coordinates. To better represent the characteristic of coronary arteries, we continue to use the coordinates of the starting point, the end point and the midpoint, add the vector between the first point and the second point, and replace the vector from the starting point to the end point with the vector from start point to the center point, and the center point to last point to better capture the shape of the artery.

\vspace{-8pt}
\subsection{Graph models}
%

We consider the following GNN methods for coronary artery labelling: (i) \textit{GCN} proposed by Kipf and Welling~\cite{kipf2017semi}, which is a spectral-based GNN with mean pooling aggregation; (ii) \textit{GAT} graph attention networks~\cite{velivckovic2017graph}, which is a spatial-GCN model incorporating masked self-attention layers in the graph convolutions, and uses a neural network to learn neighbor-specific weights; (iii) \textit{GIN} the graph isomorphism network~\cite{xu2018powerful} is a spatial-GCN that aggregates neighborhood information by summing the representations of neighboring nodes; and (iv) \textit{GraphSAGE} which is a spatial-GCN and uses node embeddings with max-pooling aggregation~\cite{hamilton2017inductive}. 

\vspace{-8pt}
\section{Experiments and results}

\vspace{-5pt}
\subsection{Dataset and evaluation metrics}

We train and test all models on a private  dataset containing 141 patients, collected from Wuhan Union Medical College Hospital between 2020 and 2022. Data is annotated by three cardiothoracic radiologists, each with more than two years of experience, in a two-stage process. In the first stage, two experts mark each branch separately. The second stage compares the results, and differences are discussed and voted upon by three experts to determine the final label. Finally, the consolidated set of annotations constitute our experimental dataset. To the best of our knowledge, a public dataset for coronary artery labeling is not presently available.

We used the same definition as ~\cite{yang2020cpr} to determine branches and segments. The min (the simplest number of vascular branches), max (the most complex situation in the dataset), and Avg (the average number of dataset) branches and segments are given in Table~~\ref{table:detail-dataset} and are higher than previous work~\cite{yang2020cpr} (22, 3, and 13.23 respectively), demonstrating the complexity of the dataset. To verify our graph representation  method, we removed the L-PDA and L-PLB branches to create an 11 category dataset as per~\cite{yang2020cpr}.

\begin{figure}[!t]
\centering
\includegraphics[width=0.8\linewidth]{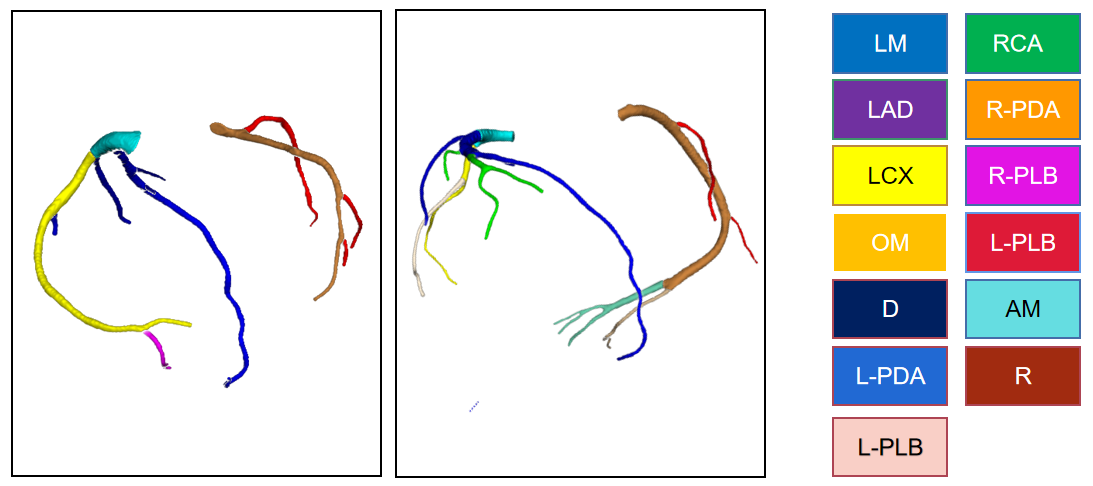}
\vspace{-7pt}
\caption{
Representations of specific segments, and the connections for two different subjects in the dataset (left and right boxes). The right vessel tree, which has 13 segments, is more complete, while the left tree lacks some vessel segments.}
\label{fig:Fig4}
\vspace{-15pt}
\end{figure}

\begin{figure*}[!t]
\centering
\includegraphics[width=0.62\linewidth]{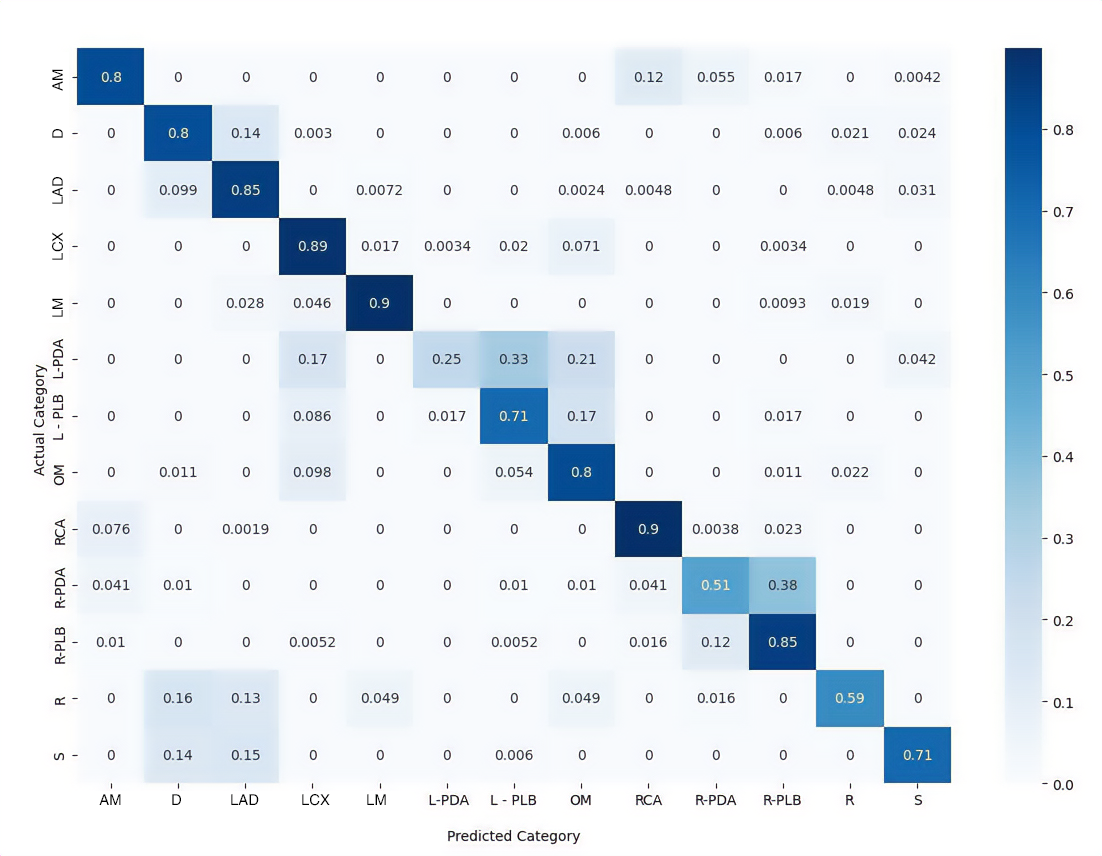}
\vspace{-6pt}
\caption{
Normalized confusion matrices for labeling segments (13 classes).
}
\label{fig:Fig6}
\vspace{-6pt}
\end{figure*}

In this work, we follow the Society of Cardiovascular Computed Tomography (SCCT) guidelines~\cite{leipsic2014scct} for anatomical labeling of the coronary arteries, represented by 13 coronary segments (classes). Previous methods have only used 9 or 11 classes ~\cite{zhou2020hybrid,hampe2021graph,yang2020cpr}. Segments are the branches after bifurcating, and graph edges capture relationships between segments.
The coronary arteries are composed of two major parts, the left and right sub-trees, both originating from the aorta. 
The left side is typically connected to the aorta by the left main artery (LM), which divides into two branches, the left anterior descending artery (LAD) and left circumflex artery (LCX). Multiple side branches originate from main branches: the ramus (R) and obtuse margin (OM) from the LAD; and the diagonal artery (D) from the LCX. Some LCX terminals will differentiate into the left posterior descending coronary artery (L-PDA) and left posterior lateral branch (L-PLB). 
The main artery of the right sub-tree is the right coronary artery (RCA), with common side branches such as Acute marginal branches (AM), right posterior lateral branches (R-PLB) and the right posterior descending artery (R-PDA) (See Fig~\ref{fig:Fig3}).
Details of our dataset are given in Table~\ref{table:detail-dataset}.

CCTA images are scaled to $v=0.5mm$ voxel spacing. After centerline extraction, we adjust the control points on the centerline to merge very close branches into the same branch, reducing unnecessary segmentation. At the same time, we resample control points such that control points are spaced at the same distance of 10 voxels as\cite{yang2020cpr}.

We use the weighted-F1 ($W\:F1$) score to measure performance as this is a multi-class classification task with a highly uneven class distribution.
This metric is defined as $W\:F1 = \sum_{i=1}^{13} \frac{2 \times precision_i \times recall_i }{precision_i + recall_i} \times w_i$,
where $w_i$ is the weight (total percentage of samples that belong to a class) of the $i-th$ class.

\vspace{-6pt}
\subsection{Experimental setup and implementation details}

Data is randomly divided into five equal sized subsets, and models are evaluated using a 5-fold cross validation. 
Graph models are implemented using Pytorch geometric~\cite{Fey/Lenssen/2019}.
To demonstrate our graph representation method, we use a model composed of two GCN layers followed by an FC layer. We use the graph structures (i.e. nodes and edges) generated by different methods as inputs to the graph networks to predict coronary artery labels.
To verify the effect of different graph neural networks on the automatic coronary artery labeling task, all graph networks (including GAT\cite{velivckovic2017graph}, GIN\cite{xu2018powerful}, Graphsage\cite{hamilton2017inductive}) use the same structure, and receive the graph generated by our method as input.
%

Models were trained for 500 epochs with a batch size of 8, with categorical cross-entropy loss and the Adam optimizer~\cite{kingma2014adam} (with a learning rate of $1e-3$). These hyper-parameters are fine-tuned based on the performance of the models on the validation set.

\vspace{-6pt}
\subsection{Results and discussion}
The performance of our proposed method for graph representation, and for coronary artery labeling using a variety of graph networks, are presented in Tables \ref{table:graph_construction} and \ref{table:graph_classificationn} respectively. 

\textit{Graph representation and embeddings: }
We compare the performance of our proposed graph representation for artery labeling (using 11 and 13 classes) with ~\cite{yang2020cpr}, using a GCN implementation as per~\cite{kipf2017semi}. For a fair comparison, we only considered the positional features of ~\cite{yang2020cpr} (global features are excluded) with our node embeddings, using the same proposed graph representation. 
Results in Table \ref{table:graph_construction} show superior performance of our proposed method over the baseline~\cite{yang2020cpr}. Using our proposed repeatable approach with extra features including directional vectors (see Section \ref{subsec:egconstruction}) achieves better characterizes and representation of coronary artery branches.

\textit{Graph network evaluations: }
We compared the performance of different graph-based learning methods using our proposed graph representation and positional embeddings. Results are presented in Table\ref{table:graph_classificationn}. 
%
For both 11 and 13 classes GraphSAGE outperforms other graph-based models. This is due to the dataset's imbalanced nature, the small number of subjects, and the training paradigm of GraphSAGE. Unlike other graph networks, in each iteration GraphSAGE runs over a sampled sub-graph in a node-wise scheme, resulting in high performance on our dataset and some benefits when faced with imbalanced data (though rare classes remain problematic).  

The confusion matrix for the 13-class evaluation obtained from the application of GraphSAGE is presented in Fig~\ref{fig:Fig6}. The low performance of GraphSAGE for R, L-PDA and R-PDA is due to the very low number of samples for these classes. For instance, only one third of subjects have R branches, and even fewer show L-PDA and R-PDA branches. We note that it is difficult even for experts to label these three branches due to their structures.

\begin{table}[!t]
\caption{Comparison of different graph embedding with the proposed graph presentation for labeling coronary artery segments with a standard GCN model.}
\vspace{-5pt}
\centering
\label{table:graph_construction}
\resizebox{0.48\textwidth}{!}{%
\begin{tabular}{c c c c c}
\toprule
\textbf{Graph representation} & \textbf{Node embeddings} & \textbf{F1-score (11)} & \textbf{F1-score (13)} \\
\midrule
Our & \cite{yang2020cpr}  & 0.644 &   0.638\\
Our & Our                 & 0.652 &  \textbf{0.654}\\
\bottomrule
\end{tabular}}
\vspace{-6pt}
\end{table}

\begin{table}[!t]
\caption{Comparison between graph models with the proposed graph construction methodology.}
\vspace{-5pt}
\centering
\label{table:graph_classificationn}
\resizebox{0.34\textwidth}{!}{%
\begin{tabular}{l c c c }
\toprule
\textbf{Graph Model} & \textbf{F1-Score (11)} & \textbf{F1-score (13)} \\
\midrule
GCN  &  0.652  & 0.622 \\
GAT  &  0.648  & 0.642 \\
GIN  &  0.664  & 0.657\\
GraphSAGE  & 0.812   & \textbf{0.805}\\
\bottomrule
\end{tabular}}
\vspace{-12pt}
\end{table}
\vspace{-8pt}
\section{Conclusion}
\vspace{-5pt}

In this study we introduced a graph-based labeling pipeline, including a robust graph representation method combined with a GraphSAGE network to label segments of coronary arteries from CCTA scans.
Promising results are obtained, and the graph structure has a high data representation efficiency and strong feature encoding capacity.
The proposed approach has broad potential applicability, with automatic coronary labeling of use to greatly expedite the process of generating the reports which form the basis of diagnosis.

\section{Compliance with Ethical Standards}
\vspace{-5pt}
The experimental procedures involving human subjects data described in this paper were approved by the Research Ethics Board of Wuhan Union Hospital. Waiver for written patient consent was sought for the data collection and analysis.


\bibliographystyle{IEEEbib}
\bibliography{refs}

\end{document}